\documentstyle[eqsecnum,aps,preprint,epsfig]{revtex}
\begin{document}
\draft
\author{D. E. Grupp$^{*,1,2}$, T. Zhang$^1$, G. J. Dolan$^1$, 
Ned S. Wingreen$^2$}
\bigskip
\address{ $^1$Department of Physics \\
David Rittenhouse Laboratory \\
University of Pennsylvania,
Philadelphia, PA 19104}
\bigskip
\address{$^2$NEC Research Institute\\
4 Independence Way\\
Princeton, NJ 08540}
\title{Observation  of a Nanoscale Metallic Dot 
Self-Consistently Coupled to a Two-Level System}
\maketitle \maketitle
\begin{abstract}
We have observed anomalous transport properties for a $50$ nm Bi dot 
in the Coulomb-blockade regime.  Over a range of gate voltages, Coulomb
blockade peaks are 
suppressed at low bias, and dramatic structure appears 
in the current at higher bias. 
We propose that the state of the dot is determined {\it self-consistently}
with the state of a nearby two-level system (TLS) to which it is 
electrostatically coupled. 
As a gate voltage is swept, the ground state
alternates between states of the TLS, leading to skipped Coulomb-blockade
peaks at low bias.
At a fixed gate voltage and high bias, transport may occur 
through a cascade of excited states connected by the dynamic 
switching of the TLS.
\end{abstract}
\vspace{.5in}
\tightenlines
With the Coulomb blockade in single-electron 
transistors\cite{fulton,CB} and quantum 
dots\cite{charlie,meir} 
now firmly established, efforts have 
begun to turn towards more complex systems in which these 
artificial atoms form the building blocks.
Significant progress has been made in understanding 
linear-response transport in what may be 
termed ``artificial molecules" in studies of double quantum 
dots\cite{waugh,matveev},  as well as ``artificial solids" 
in studies of arrays of quantum 
dots\cite{leo,stafford}. Moreover, such multi-dot systems have been considered
as the basis for novel computer memory and logic elements\cite{Lent}. 
In these applications the transport properties must be considered 
well beyond the regime of linear-response.

In this Letter, we examine the transport 
properties of a quantum dot coupled to a single two-level system (TLS)
as a model for nonlinear transport phenomena in systems
containing artificial atoms.
We report a set of anomalous transport properties of a small Bi dot
in the Coulomb-blockade regime. The anomalies are explained by the
presence of a TLS in close proximity to the dot. 
While TLS's have been studied as a source of $1/f$ noise in 
MOSFET's\cite{RTS:Kogan,RTS:MOS},  
quantum point contacts\cite{RTS:Kogan,RTS:QPC}, 
and single-electron transistors (SET's)\cite{RTS:SET,zimmerman}, 
exhibited as random telegraph signals, 
the coupling in our devices leads to {\it non-stochastic} configurations 
of the TLS.
At low bias, reversible switching of the TLS leads to skipped Coulomb
blockade peaks in a manner similar to peak suppression in double 
quantum dots\cite{glazman}.
At higher bias, a novel effect emerges in which 
transport occurs through a cascade of excited states 
which are connected by the {\it dynamic} switching of the TLS.
Similar effects may be expected in multi-dot systems.

The devices, shown schematically in the inset to Fig. 1, are SET's
in the standard double-junction geometry with a
capacitively
coupled gate\cite{fulton,CB}.
They were fabricated using e-beam lithography to make a 
shadow-mask for use
with the self-aligned double-angle evaporation technique\cite{fab}.
The leads, consisting of Cu wires $20$ nm  in diameter,
were deposited first. They were then oxidized {\it in-situ} for 30 minutes
at an oxygen pressure of $1\times 10^{-4}$ Torr before deposition 
of the Bi dot. The resultant junction resistances of $R_J\geq 1$~M$\Omega$
satisfy the condition for the Coulomb blockade $R_J > R_q=
h/4e^2 \simeq 6.5 {\rm k}\Omega$. 
Voltage-biased DC measurements were performed in an 
Oxford dilution refrigerator held at the base
temperature of 70 mK.

In the measured devices, 
the diameter of the dot is roughly $50$~nm. As the size of the
Bi dot is
smaller than a typical grain in a film of Bi grown on the same
surface, the dot is
likely   a single crystal.  Furthermore,  while 
quench-condensed amorphous Bi is superconducting, crystalline Bi
is not.
Bulk crystalline Bi is a semimetal, with a low Fermi energy, $E_F=27.2$~meV,
and a low
electron density, $n=2.7\times 10^{17}/{\rm
cm^3}$
\cite{issi}, about $10^5$ times lower than a typical metal. A small
overlap of
bands at the Fermi level leads to an equal number of electrons and
holes.
The highly anisotropic
Fermi surface results in Fermi wavelengths
$\lambda _F=$14--215~nm. These wavelengths are  comparable to
the dimensions
of the dot, and quantum properties might be 
expected\cite{charlie,halp}. However, the devices reported here
exhibited only metallic behavior.

Experiments were also conducted using Al as the dot material.
The five Al devices studied exhibited only the standard 
Coulomb-blockade behavior.
The anomalous behavior presented here
occurred in all seventeen of the Bi dots studied.

The most striking feature of the data is seen in $IV$ curves presented
as a function of both bias and gate voltages, $V_b$ and $V_g$, respectively.
In a series of $IV_b$ curves taken over a full period of the Coulomb blockade
in $V_g$, {\it an anomalous gap remains} where the blockade would normally
vanish (Fig. 1). This can also be seen in
$IV_g$ sweeps in the same region (Fig. 2(a)), where 
at low bias ($V_b\ll E_c/e$, where $E_c$ is the charging energy of the dot),
the conductance peaks disappear. 
The range of gate voltages in which the suppression occurred
varied from device to device, as 
did the number of missing peaks, with a maximum of 
about five. 
Each device exhibited only one region of peak suppression within 
the range of gate voltages studied, which corresponded to changing the 
number of electrons on the dot by up to 100.  

At higher biases, where $V_b$ is greater than the anomalous gap, but 
still smaller than $E_c/e$,  the missing 
peaks emerge and typically split, as seen in Fig. 2(a). At even
higher biases, the peaks are suppressed
within an envelope, and the lineshapes become complex.  
An example of this from a different device is shown in Fig. 3.  A 
notable feature of this device is the noise in the peaks, 
visible even at the measurement bandwidth of 1 Hz.
The noise has a broad spectrum, as evidenced in measurements
with a bandwidth of 1 kHz (Fig. 3, left inset).

Before introducing the TLS, we first consider other possible
origins of these anomalies.
A gap similar in appearance to Fig. 1 may arise in 
superconductors\cite{scgap}. This would require that both
the dot and the leads are superconducting. While Bi in its 
amorphous state may be superconducting, the Cu leads are not.
Moreover, a magnetic field up to 9 T does not remove the 
anomalous gaps.  The opening 
of a gap at the band overlap in the Bi
semimetal due to size quantization may also be
considered\cite{lut}, but such a bandgap would not manifest itself as in 
Fig. 1.
The result of a bandgap $E_g$ in the Bi dot would be to 
increase the spacing in gate voltage
$\Delta V_g$ between the conductance peaks which occur at the band
edges\cite{meir}:
\begin{equation}
\Delta V_g=\frac{C_\Sigma}{eC_g} (E_c+E_g)  
= \frac{e}{C_g} + \frac{C_\Sigma}{C_g} \frac{E_g}{e},
\label{DeltaV}
\end{equation}
where $E_c = e^2/C_{\Sigma}$ is the charging energy, and 
$C_g$ and $C_\Sigma$ are, respectively, the capacitance of
the dot to the gate and the total capacitance of the dot.  
Finally, we may exclude migrating charges
as the data are highly reproducible.

In what follows, we show that the transport anomalies may be
explained by postulating a TLS coupled electrostatically to
the Bi dot.  We imagine the TLS to consist of a double-well potential 
(Fig. 1 inset) in
which a charged particle is free to move.  
Electrons on the dot interact
with the charge in the TLS. Hence a change in the number of electrons
on the dot may change the relative energies of the two states of the TLS. 
In turn, the switching  of the TLS will act 
as an effective change in gate voltage for the dot. Thus, the
charge state of the dot  must be determined {\it self-consistently} with the
state of the TLS. 

To model in  detail the behavior of a dot electrostatically
coupled to a two-level system, we consider the following 
classical Hamiltonian:
\begin{equation}
H^{\scriptscriptstyle \rm ON/OFF} = {1 \over {2 C_\Sigma} }  
(Q - Q_0 - Q_0^{\scriptscriptstyle \rm ON/OFF})^2
+ V_g Q_{\rm TLS}^{\scriptscriptstyle \rm ON/OFF} 
+ E_{\rm TLS}^{\scriptscriptstyle \rm ON/OFF},
\label{H}
\end{equation}
where ON/OFF denotes the state of the TLS.
Without the terms due to the two-level system, this
is just the ordinary Hamiltonian of a metal dot. The 
actual integer charge on the dot is given by $Q$, while 
the optimal charge is controlled by the gate voltage 
$Q_0 \propto V_g C_g$. The electrostatic
coupling between the TLS and the dot can be expressed 
by a shift $Q_0^{\scriptscriptstyle \rm ON/OFF}$ 
of the optimal charge on the dot.
The TLS is also electrostatically coupled to the gate, yielding
the linear term $V_g Q_{\rm TLS}^{\scriptscriptstyle \rm ON/OFF}$. 
Lastly, there  may be an internal 
energy difference between  the states of the TLS, represented by 
$E_{\rm TLS}^{\scriptscriptstyle \rm ON/OFF}$. 
The model is also applicable to multilevel systems,
in which case the TLS parameters can 
take on several discrete values.

By an appropriate choice of $Q_0$, we may set the 
OFF-state parameters to be zero.
This choice for the Hamiltonian leads to the two families of 
parabolas shown in Fig. 2(b). 
The horizontal offset, vertical offset, and slope of the 
ON-state parabolas are determined by $Q_0^{\scriptscriptstyle \rm ON}$,
$E_{\rm TLS}^{\scriptscriptstyle \rm ON}$, 
and $Q_{\scriptscriptstyle \rm TLS}^{\scriptscriptstyle \rm ON}$, 
respectively.  

As shown in Fig. 2(b), as the gate voltage 
is increased, the lowest energy configuration may switch back
and forth many times between the TLS OFF 
and ON states. This 
provides a natural explanation for the missing 
Coulomb-blockade peaks at low bias voltages in Fig. 2(a).
For example, the first switching ON  of the
TLS occurs together with a switch from $N$ to $N+1$ 
electrons on the dot (first dark circle in Fig 2(b)). 
For this double switch to occur, either the electron number 
must first change with the TLS fixed, or the TLS must first switch with 
the electron number fixed. Either of these processes alone requires
activation energy. Hence the double switch has an
{\it activation barrier} and is too slow to produce a 
measurable current \cite{hyst}. This process preempts the usual
Coulomb-blockade peak which would have occurred at slightly higher 
gate voltage. The peaks continue to be suppressed at low bias 
until increasing $V_g$
causes the TLS to switch permanently ON,
at which point conductance peaks resume.
The number of missing peaks can be estimated from Eq.~(\ref{H})
as\cite{qzero}
\begin{equation}
\label{eq:numpks}
\frac{e}{|Q_{\rm TLS}^{\scriptscriptstyle \rm ON}|}
{\rm Mod}\left(\frac{|Q_0^{\scriptscriptstyle \rm ON}|}{e}\right)
[1 - {\rm Mod}\left(\frac{|Q_0^{\scriptscriptstyle \rm ON}|}{e}\right)],
\end{equation}
such that a small slope $|Q_{\rm TLS}^{\scriptscriptstyle \rm ON}|$ 
can lead to a large number of missing peaks.   

To understand the more complex behavior at higher bias voltages, we have
simulated transport through the dot using the orthodox 
Coulomb-blockade model\cite{CB}, but including the TLS. 
We consider the transport 
current between a right lead with voltage $V_{b}/2$ and a 
left lead with voltage $-V_b/2$, both leads coupled via equal junction 
resistances $R$ to the dot\cite{footnote}. 
Assuming rapid thermalization
of electrons on the dot, we solve the rate equations for transitions between 
different charge states of the dot.
However, we also allow transitions to occur between the two states of the 
TLS. The transition rate is $\Gamma_{\rm TLS}$ 
if the overall energy is thereby
lowered, and $\Gamma_{\rm TLS}\,\exp(-\Delta 
E/k_BT)$
if the overall energy is raised by $\Delta E$. 

As shown in Figs. 2(a) and 3, the simulated current is 
similar to that observed experimentally, remarkably so in the
latter. 
In both cases, the fitting parameters are those of the TLS (Eq.~\ref{H}).
In Fig. 2(a), the best fit was obtained in the 
saturated limit of $\Gamma_{TLS}$ much faster than the electron 
tunneling rate, but the opposite limit was obtained in Fig. 3. 
Thus there appears to be a considerable variation 
of TLS tunneling rates from device to device. 

The interactions between the dot and the TLS
may lead to a novel transport process in which the TLS is 
{\it active}, switching dynamically. 
Consider the inset of Fig. 2(b), which shows the states of the 
system and the simulated current. There are four accessible states
{\bf A}-{\bf D} ($V_b>0.2E_c$), with $N$ or $N+1$ electrons on the dot, 
and with the TLS either ON or OFF. 
At this gate voltage, the bias voltage is sufficient to add an electron
to the dot,
going from state {\bf A} to {\bf B}, with the TLS OFF.
The extra electron may escape into the other lead,
causing relaxation from {\bf B} back to {\bf A}.
However, it is also 
possible for the TLS to spontaneously switch from OFF to ON, since $\Delta E$
is negative for this process, thereby taking the system from
state {\bf B} to state {\bf C}.  
Now, the system may switch repeatedly 
between states {\bf C} and {\bf D} resulting
in electron transport, 
with the TLS staying in the ON state.  Alternatively, when the system is in
state {\bf D} with $N$ electrons, 
the TLS may switch back to OFF without changing 
the number of electrons on the dot, and the system is back where 
it started in state
{\bf A}. This process can then begin again, each cycle transferring a
minimum of two electrons through the dot.

If the TLS transition rate $\Gamma_{\rm TLS}$ is large compared
with the tunnelling rate, then these cycles, where each succesive electron
moving through the dot is accompanied by a switch of the TLS, will
dominate the transport. If the TLS is slow, then the 
system will switch telegraphically back and forth between 
transport with the TLS ON or OFF, 
each process exhibiting a different conductance. This may explain the noise
seen in the device of Fig. 3. 

While it is evident that the TLS is associated with Bi, since these 
effects were not observed in our Al dots,
the microscopic origin of the
TLS is unknown. It may reside on the surface of the dot in the Bi
oxide layer, or it may be due to an ancillary grain of Bi from the
double-angle evaporation. 
If it were a bistable defect, the associated small charge displacement,
on the order of a lattice spacing,
would be an insignificant effect in our device geometry.
A larger effective charge displacement may come from a 
trapping state which is either filled or unfilled\cite{zimmerman}. 
The dynamics of such a charge trap may then be 
responsible for the broad range of lifetimes observed.

The phenomenology of the dot-TLS system may be important in 
a number of applications.
In quantum dots,
a system consisting of two dots connected in parallel\cite{doublepar}
will mimic a dot coupled to a two-level system if one of the dots is
electrically coupled to only one lead. 
With perhaps broader implications,  a single-electron transistor used as an 
electrometer\cite{electrom} may behave in a way similar
to the dot-TLS system.  Such an electrometer, 
which may be used to detect  charge configurations, 
{\it interacts with the system being measured}.  
For example, the 
Bi dot in our devices may be said to measure the state
of the TLS, but clearly this process sometimes changes the
state of the TLS.  Therefore, in using an electrometer to probe
a system of mobile charges, additional movement of the charges
due to the measurement process must also be considered.

In conclusion, we have observed anomalous transport characteristics in a
$50~{\rm nm}$ Bi dot, including suppression of Coulomb-blockade 
conductance peaks at low bias voltages, and splitting of these peaks at higher
bias.
This behavior may be ascribed to an electrostatic  coupling 
of the dot to a TLS. 
The interaction with the dot causes the 
TLS to switch back and forth multiple times as the gate voltage is
increased. 
Furthermore, sufficient bias may result in a transport process where
the TLS switches dynamically with transport of each electron through the dot.
A similar effect is
expected in some double-dot systems. 
It may be informative to study the 
dynamics of such dot-TLS systems, particularly if the switching rate
of the TLS can be controlled. Finally, more complicated dynamics
may be expected in dots coupled to multilevel systems.


\begin{figure}
\caption{$I$ vs. $V_b$ curves as a function of gate
voltage $V_g$, over a full period of
the Coulomb blockade corresponding to the region 
of peak marked with an arrow in Fig. 2(a). An anomalous gap
occurs where the current is
expected to be nearly linear, as
indicated by the dotted line.
Inset: Schematic of the devices, consisting of a Bi dot coupled
to two Cu leads via tunnel junctions, with a capacitively  coupled gate. 
Nearby is the inferred  two-level system (TLS) 
containing a single charge which is electrostatically
coupled to the dot. 
}
\end{figure}
\begin{figure}
\caption{
A series of $I$ vs. $V_g$ sweeps in (a) for  bias
voltages $V_b>0$, $T=70\, {\rm mK}$.
Offset for
clarity, the flat regions mark $I=0$ for each curve. 
The top curve is the 
simulated current, for
$V_b = 4\, {\rm mV}, E_c/e=8.2\, {\rm mV}$ 
and $T = 70\, {\rm mK}$ as in
the top experimental curve, and fitted two-level system (TLS) 
parameters (Eq.~\ref{H})
$Q_0^{\scriptscriptstyle \rm ON}/e = 0.23$,
$Q_{\rm TLS}^{\scriptscriptstyle \rm ON}/e = -0.1$, 
$E_{\rm TLS}^{\scriptscriptstyle \rm ON}/E_c = 0.39$,
and $\Gamma_{\rm TLS} = 100 hE_c/e^2R$.
The rounding of the data relative to the
simulation suggests that the temperature of the dot
electrons is higher than $70\, {\rm mK}$.
The energy of the dot-plus-TLS system is shown in (b) 
for both states of the TLS (solid line=OFF, dashed line=ON). 
$N$ indicates the number of electrons on the dot.
For low bias
$V_b \ll E_c/e$, the system will follow the lowest energy curve,
switching at the crossover points indicated (dark circles), 
preempting the crossovers associated with Coulomb-blockade peaks.
Inset:  
Expanded view of the states relevant to the peak marked with an arrow in (a), 
shown with the simulated current.
At higher bias, transport will take place through the sequence
of states {\bf A}-{\bf D} where the TLS switches with each electron
passing through the dot,
as described in the text.
}
\end{figure}
\begin{figure}
\caption{
Effect of the TLS switching at a bias $V_b\simeq E_c/e$
for a different device than in Fig. 2 (solid line).
The simulated current (dotted line) reproduces many features of
the data with fitted
TLS parameters 
$Q_0^{\scriptscriptstyle \rm ON}/e =-0.3$,
$Q_{\rm TLS}^{\scriptscriptstyle \rm ON}/e = -0.048$, 
$E_{\rm TLS}^{\scriptscriptstyle \rm ON}/e /E_c = 0.27$, 
and $\Gamma_{\rm TLS}^{\scriptscriptstyle \rm ON}/e = 0.01 hE_c/e^2R$,
and the experimental parameters
$V_b = 0.9\, {\rm mV}, E_c/e=1.5\, {\rm mV}$, $T = 140\, {\rm mK}$, and 
bandwidth=1 Hz.
The temperature used in the fit was $T=345$~mK, indicating that the small,
isolated grain of Bi may not cool to the temperature of the leads.
Left inset: Expanded view of peak indicated by arrow, with bandwidth=1 kHz.
Right inset: Simulated current shown by itself for clarity. }
\end{figure}

\pagebreak

\begin{figure}
\centerline{\epsfig{figure=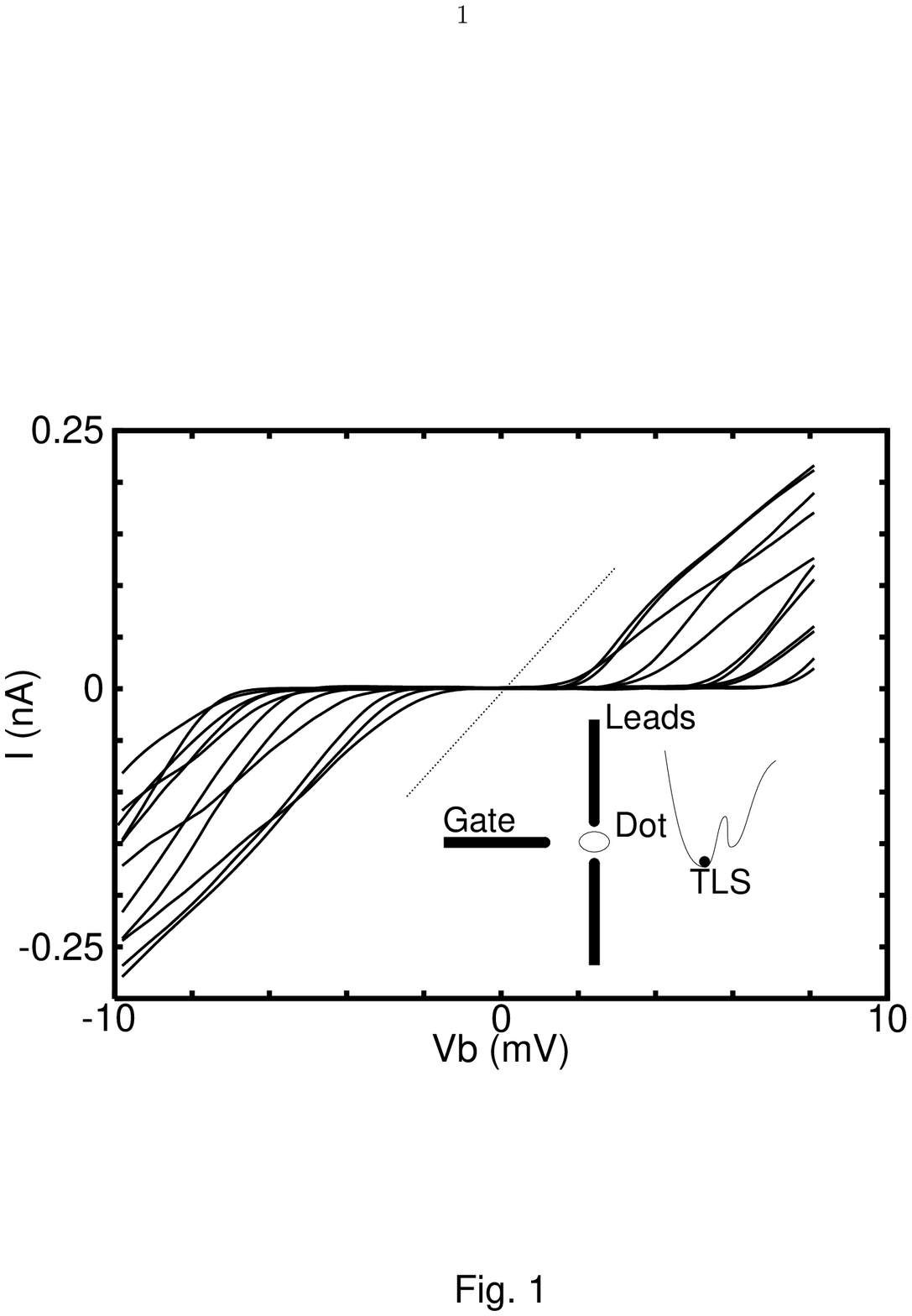,height=6.0in}}
\label{}
\end{figure}
 
\pagebreak
\begin{figure}
\centerline{\epsfig{figure=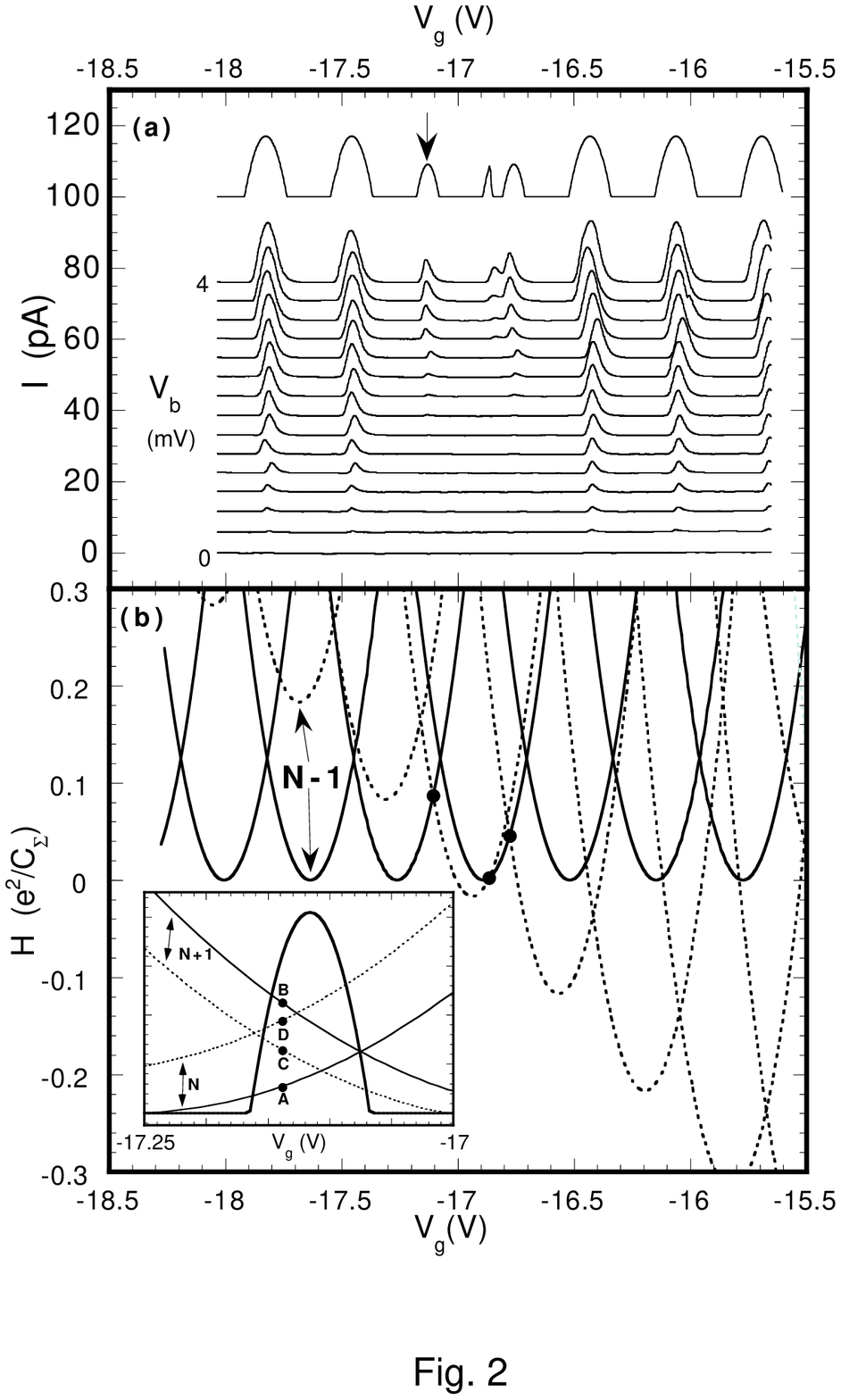,height=8.0in}}
\label{}
\end{figure} 

\pagebreak
\begin{figure}
\centerline{\epsfig{figure=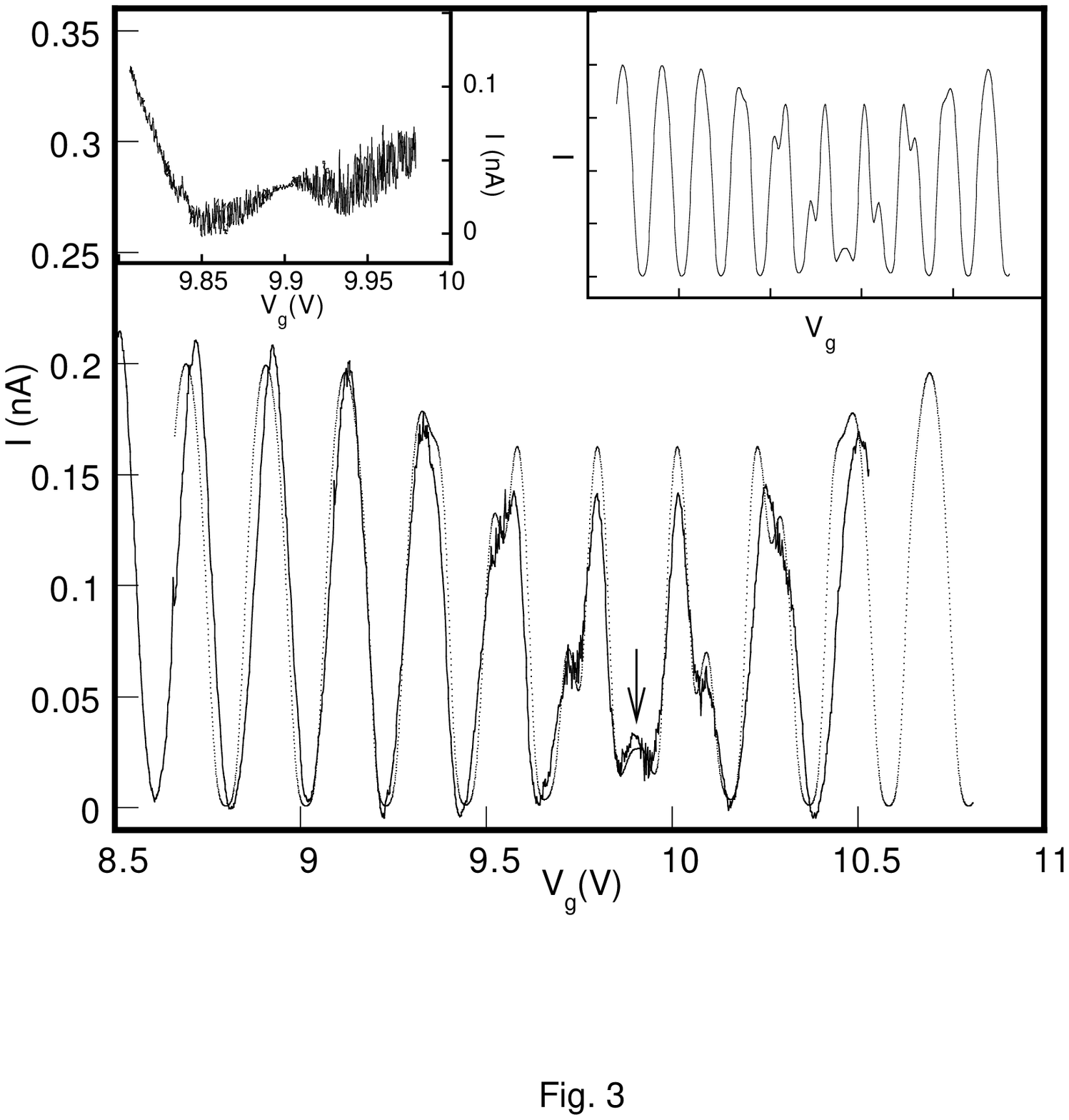,height=7.0in}}
\label{}
\end{figure} 


\begin{thebibliography}{99}
\bibitem{hyst}
The double switch of the TLS and the charge state of the dot 
is fast enough that no hysteresis is observed in the experiment
as the gate voltage is swept. This implies a rate for the double 
switch faster than a few Hertz, the bandwidth of the measurements.
\bibitem{qzero}
$|Q_0^{\scriptscriptstyle \rm ON}|$ will only exceed 1 if the mobile
charge of the TLS is more than one electron.
\bibitem{footnote}
For simplicity, we assume zero
capacitance and symmetric resistances 
between the leads and the dot, though finite capacitances
and asymmetric resistances 
can be simply included within the orthodox model of the Coulomb blockade.
\end{thebibliography}
\end{document}